\documentclass[a4paper,10pt]{article}
\usepackage[utf8]{inputenc}

\usepackage{publication}
\usepackage[backend=biber,sorting=none]{biblatex}
\usepackage{amsmath} 
\usepackage{graphicx}
\usepackage{amsfonts}

\title{Implications of temporal sampling in voltage imaging microscopy}

\author[1]{Jakub Czuchnowski}
\contact{Correspodning author: Jakub Czuchnowski - jczuchno@bu.edu}
\affil[1]{Department of Biomedical Engineering, Boston University, Boston, MA, 02215, USA}
\affil[2]{Photonics Center, Boston University, Boston, MA, 02215, USA}
\author[1,2]{Jerome Mertz}

\date{\today}

\begin{document}

\maketitle

\begin{abstract}
\textbf{Significance:} Voltage imaging microscopy has emerged as a powerful tool to investigate neural activity both \textit{in vivo} and \textit{in vitro}. Various imaging approaches have been developed, including point-scanning, line-scanning and wide-field microscopes, however the effects of their different temporal sampling methods on signal fidelity have not yet been fully investigated.
\\
\textbf{Aim:} To provide an analysis of the inherent advantages and disadvantages of temporal sampling in scanning and wide-field microscopes and their effect on the fidelity of voltage spike detection.
\\
\textbf{Approach:} We develop a mathematical framework based on a mixture of analytical modeling and computer simulations with Monte-Carlo approaches. 
\\
\textbf{Results:} Scanning microscopes outperform wide-field microscopes in low signal-to-noise conditions and when only a small subset of spikes needs to be detected.  Wide-field microscopes outperform scanning microscopes when the measurement is temporally undersampled and a large fraction of the spikes needs to be detected. Both modalities converge in performance as sampling increases and the frame rate reaches the decay rate of the voltage indicator.  
\\
\textbf{Conclusions:} Our work provides guidance for the selection of optimal temporal sampling parameters for voltage imaging. Most importantly it advises against using scanning voltage imaging microscopes at frame rates below 500 Hz. 
\end{abstract}



\begin{multicols}{2}

\section{Introduction}
Voltage imaging microscopy has emerged as a powerful tool to investigate neural activity both \textit{in vivo} and \textit{in vitro} \cite{peterka2011imaging,xu2017voltage}. Multiple approaches have been developed, including point-scanning \cite{weber2023high,akemann2013two,platisa2023high} , line-scanning \cite{xiao2024large} and wide-field\cite{xiao2021large,tominaga2023stable,lu2023widefield} microscopes. However, differences in their method of temporal sampling and their effect on the recorded voltage traces have not been properly studied. Additionally, because of  variations in their design architectures, the experimental comparison of these systems is generally not straightforward. This often leads to choices of which system to use based on imperfect heuristics, rules of thumb or intuition, rather than concrete  quantitative metrics. In this work, we aim to develop a theoretical understanding of how the choice of different microscope systems  (and therefore different temporal sampling approaches) affects the shape and fidelity of the recorded signals in a way that is generalized and applicable to different modalities. This work focuses on spike detection as current technical limitations, in terms of imaging speed, are mainly connected to this particular use of voltage imaging microscopes (as opposed to subtreshold activity measurement).
\section{Theory}
\begin{figure}
\begin{center}
\includegraphics[width=15cm]{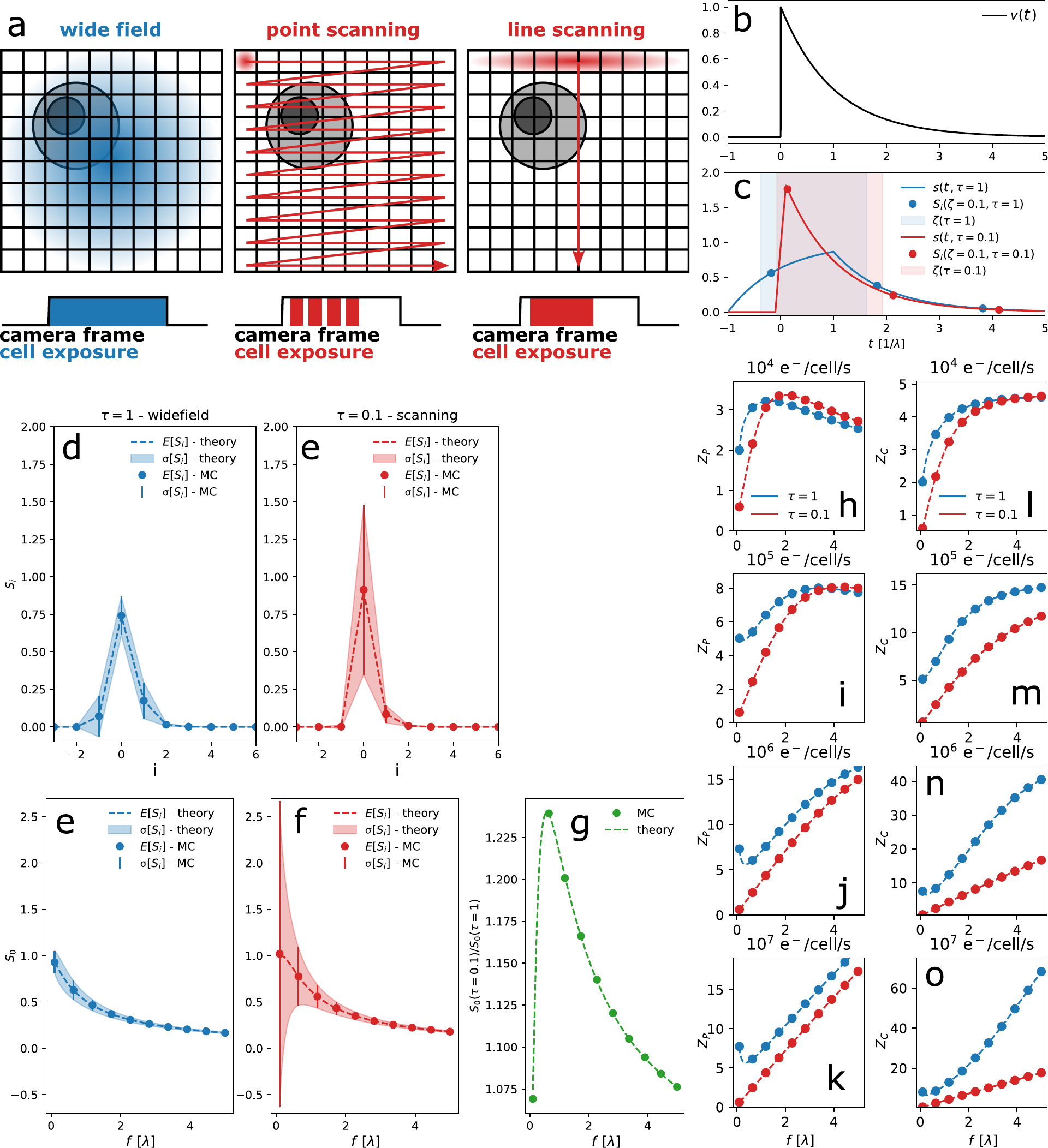} 
\caption 
{ \label{fig:1}
(a) Different imaging configurations result in different temporal sampling of cells within a camera frame. (b) Shape of the measured voltage transient. (c) Sampling functions for wide-field  ($\tau=1$) and scanning ($\tau=0.1$) configurations. (d-e) Averaged spike shapes for wide-field (d) and scanning (e) configurations ($f=0.4$) where $\sigma[X] = \sqrt{V[X]}$ is the standard deviation. (e-f) Dependence of the expected value of the voltage signal peak on sampling frequency (e - wide-field, f - line-scan).  (g) Ratio of the expected peak values between scanning and wide-field configurations. (h-k) Z-score comparison of peak detection (PD) fidelity between wide-field and scanning configurations. (l-o) Z-score comparison of template matching (TM) fidelity between wide-field and scanning configurations for different numbers of detected photons.} 
\end{center}
\end{figure}

Different imaging configurations result in different temporal sampling of cells within a frame (\textbf{Fig \ref{fig:1}a}). Because we focus here on the temporal aspects of different microscope modalities, we do not consider spatial aspects and assume that the analyzed voltage traces come from averaging over the cell image. To describe the filtering effects of a finite integration time on the measurement of a continuous voltage indicator signal  $v(t')$ we define a sampling function $s(t)$:

\begin{equation}
    s(t)=(v \otimes \Pi_\tau)(t) = \frac{1}{\tau} \int \limits_{t-\frac{d\tau}{2}}^{t+\frac{d\tau}{2}} v(t')dt'
\end{equation}
where
\begin{equation}
    \Pi_\tau(t)=
    \left\{\begin{smallmatrix}
       1/\tau & \text{for } |t| \leq d\tau/2 \\
       0 & \text{for } |t| > d\tau/2
    \end{smallmatrix}\right.
\end{equation}
is the integration function, $d$ is the frame duration and $\tau$ is the duty cycle (i.e. the  fraction of the frame time when the signal is being integrated). The sampling function $s(t)$ allows us to extract a particular realization of an experimental trace of the voltage signal $S_i(\zeta)=s(di+\zeta)$, where $i$ is the frame number ($i \in \mathbb{N}$) and $\zeta$ is a random offset between the voltage signal start time and the microscope frame center time ($\zeta \in  [\zeta_0,\zeta_0+d]$, where $\zeta_0$ is chosen so that all the measured voltage traces ($S_i$) are aligned in time and $S_0$ corresponds to the max detected spike height -- see Appendix 1 for an explicit form of $\zeta_0$). In this work we assume that the voltage sensor response to a spike has an exponential shape (Fig \ref{fig:1}b, which is a slight simplification of the bi-exponential shapes from experimental measurements of fluorescent voltage indicators such as Voltron \cite{abdelfattah2023sensitivity}):

\begin{equation}
    v(t)=
    \left\{\begin{smallmatrix}
       0 & \text{for } t<0 \\
       \lambda e^{-\lambda t} & \text{for } t \geq 0
    \end{smallmatrix}\right.
\end{equation}
where $\lambda$ is the relaxation rate for the voltage indicator (close to 1 ms$^{-1}$ for most voltage indicators \cite{abdelfattah2023sensitivity}). For the remainder of this work we refer to these exponential transients as 'spikes' for clarity (as they are indicative of the underlying voltage spikes) even though their shape differs from the actual shape of the membrane voltage modulation. In this case the sampling function can be calculated analytically:

\begin{equation}
    s(t)=
    \left\{\begin{smallmatrix}
       0 & \text{for } t \leq -\frac{d\tau}{2}\\
       \frac{1}{\tau}[1-e^{-\lambda(t+d\tau/2)}] & \text{for } -\frac{d\tau}{2} < t \leq \frac{d\tau}{2} \\
       \frac{2}{\tau} e^{-\lambda t} \sinh{(\lambda d \tau  /2)} & \text{for } t > \frac{d\tau}{2}\\
    \end{smallmatrix}\right.
\end{equation}

The key parameter differentiating wide-field from scanning modalities is the duty cycle $\tau$. For wide-field $\tau \approx1$ since the cell signal is integrated over almost the full frame duration; for scanning modalities $\tau<1$ due to the limited  pixel dwell time and is roughly proportional to length of the cell along the slow-axis of the microscope divided by the slow-axis width of the field of view (we use the value of $\tau=0.1$ to simulate scanning applications, as this is a reasonable value from the experimental point of view). Because the results for point scanning and line scanning microscopes quickly converge (see Appendix 2 for details) we analyze them jointly as scanning microscopes. As can be observed from Fig \ref{fig:1}c, the shape of the sampling function differs drastically between a scanning and wide-field case, with the scanning sampling function much more closely resembling the original spike shape. 

Based on the sampling function, we analytically derive both the expected value

\begin{equation}
    E[S_i(\zeta)]=\frac{1}{d} \int_{\zeta_0}^{\zeta_0+d} s(\zeta+di) d\zeta
\end{equation} 
and the variance

\begin{equation}
    V[S_i(\zeta)]=\frac{1}{d} \int_{\zeta_0}^{\zeta_0+d} s(\zeta+di)^2 d\zeta - E[S_i(\zeta)]^2
\end{equation} 
of the recorded signals from cell populations (the explicit forms for these are included in Appendix 1), as shown in Fig \ref{fig:1}d,e. Again, the averaged spike shape for the scanning case more closely preserves the asymmetry of the original spike shape, while for the wide-field case it is more symmetric owing to $E[S_{-1}(\tau=1)]\gg 0$ (a similar phenomenon is observed in experimental data obtained from \textit{in vivo} measurements \cite{abdelfattah2023sensitivity}). The spike amplitude ($S_0$) can also be calculated as a function of the normalized sampling frequency ($f=1/d\lambda$) (Fig \ref{fig:1}e,f). Normalization of the sampling frequency allows decoupling of the specific kinetic constants of the voltage indicators from our analysis; however, since $\lambda \approx1$ ms$^{-1}$ for most voltage indicators, we can consider $f$ to be approximately in units of kHz for reference. 

We observe that in the scanning case there is a dramatic increase in the variance of the recorded spike population when the normalized sampling frequency approaches low values ($f \rightarrow 0$), leading to spike amplitudes being either greatly enhanced or missed altogether (Fig \ref{fig:1}f). A similar effect does not occur in the wide-field case, suggesting that a wide-field configuration is more robust against undersampling (Fig \ref{fig:1}e). On the other hand, a scanning configuration leads to larger recorded spike amplitudes, on average (Fig \ref{fig:1}g), which in certain cases can be advantageous.

The fidelity of voltage spike detection can be estimated under different conditions. Several procedures are available for voltage spike detection, out of which we analyze the two most common: peak detection (PD) via thresholding (as this is an initial step for all of the more complex methods) and template matching (TM) \cite{kim2007automatic,pouzat2002using,franke2015bayes,rutishauser2006online,friedman1968detection,sato2007fast}. In the latter case, we choose covariance-based template matching \cite{kim2007automatic} as representative, since this has been shown to be a highly robust \cite{franke2015bayes}, while also being amenable to analytical analysis.

Peak detection (PD) is based on the assumption that a sample with a value significantly higher than the baseline noise has a high likelihood of being indicative of a spike. The fidelity of this method can thus be described by a Z-score \cite{kreyszig1979advanced} (Fig \ref{fig:1}h-k):

\begin{equation}
    Z_P=\frac{E[S_0(\zeta)]}{\sqrt{V[S_0(\zeta)]+2\sigma_N^2}}
\end{equation}
where $E[S_0(\zeta)]$ and $V[S_0(\zeta)]$ are respectively the expected value and variance of the peak measurements, and $\sigma_N$ is the standard deviation of the noise. Template matching (TM), on the other hand, calculates a similarity (in our case using correlation) $C$ between the signal $S_i$ and a template $T_i$ that corresponds to the expected spike shape \cite{kim2007automatic} :

\begin{equation}
    C(\zeta)=\sum_i T_i(S_i(\zeta)+\gamma_i)
\end{equation}
where  $T_i \equiv E[S_i(\zeta)]$, $E[\gamma_i]=0$, $V[\gamma_i]=\sigma_N^2$. Similarly to the peak detection method, we can define a fidelity metric for the template matching approach ($Z_C$, Fig \ref{fig:1}h-k):

\begin{equation}
    Z_C=\frac{E[C(\zeta)]}{\sqrt{V[C(\zeta)]+V[\sum_i T_i \gamma_i]}}
\end{equation}
where $E[C(\zeta)]=\sum_i T_i^2$, $V[C(\zeta)]=\sigma_N^2 \sum_i T_i^2+\sum_{i,j} T_iT_j Cv[S_i(\zeta),S_j(\zeta)]$ and $V[\sum_i T_i \gamma_i]=\sigma_N^2 \sum_i T_i^2$.

\section{Results}

This approach allows us to quantitatively model the expected behavior of the microscope based on the noise performance of the detector used. Assuming the voltage signal is small compared to the baseline, and, correspondingly, the signal shot noise variance is small compared to the baseline shot noise variance ($\sigma_{sn}^2=\sigma_{s}^2+\sigma_{b}^2 \approx \sigma_{b}^2$), we can write the measurement noise as:

\begin{equation}
    \sigma_n^2=\frac{\sigma_{d}^2N_{px}+\sigma_{b}^2}{f}+\sigma_{r}^2N_{px}
    \label{eq:noise}
\end{equation}
where $\sigma_{d}$ is the standard deviation of the detector dark noise, $\sigma_{b}$ is the standard deviation of the shot noise of the baseline, $\sigma_{r}$ is the standard deviation of the detector readout noise, and $N_{px}$ is the number of pixels over which the cell signal is integrated. From Eq. \ref{eq:noise} we observe that the noise variance is hyperbolically dependent on the normalized sampling frequency ($f$). As an example, for a Teledyne Kinetix camera often used for voltage imaging ($\sigma_{r}^2=1.2\ e^-$, $\sigma_{d}^2=1.03\ e^-/px/s$), a reasonable spatial sampling of a neuron ($N_{px}=100$), and a typical relative response of the voltage indicator ($\sim 10\%$),  the performance of a wide-field versus line-scan setup strongly depends on the total number of photo-electrons collected from a cell (Fig \ref{fig:1}h-o). At low signal levels ($10^4\ e^-/cell/s$, corresponding to low signal-to-noise ratio (SNR)), the spike enhancement provided by a line-scan microscope enables it to outperform the wide-field; however, for higher signal levels   ($10^6\ e^-/cell/s$), spike skipping undermines the line-scan microscope, enabling a wide-field microscope to outperform line scan.

\begin{figure}
\begin{center}
\includegraphics[width=12cm]{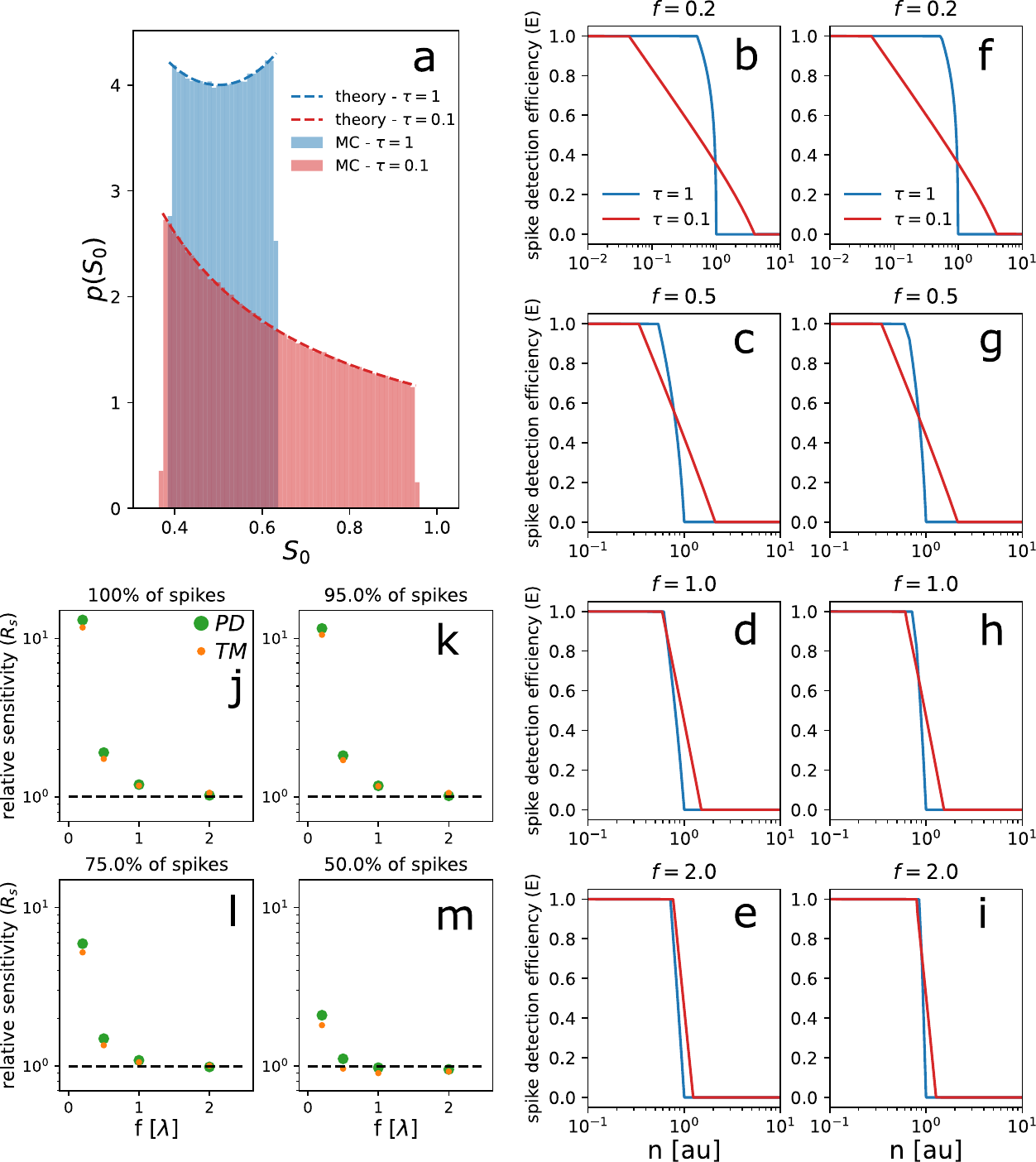} 
\caption{\label{fig:2}
(a) Probability distributions of spike heights calculated both analytically ($f=1$) and by Monte Carlo (MC). (b-i) Comparison of spike detection efficiency for both scanning and wide-field systems using PD (b-e) and TM  (f-i). (j-m) Comparison of the relative sensitivity (wide-field/scanning) for both PD and TM for different sampling frequencies.}
\end{center}
\end{figure}

Unfortunately, while Z-score metrics are powerful they do present disadvantages for our analysis: 1) they are dependent on a realistic approximation of noise, which in practical applications can be difficult to obtain, 2) they describe an average (in a statistical sense) performance of the system, which depending on the actual shape of the population distribution of the recorded spikes might not be informative to the experimenter (for example, depending on the distribution shape, the average can be very different from the median), 3) they assume a near-Gaussian population distribution, which is generally not valid since spike distributions exhibit compact support with sharp edges (Fig \ref{fig:2}a) that can lead to overestimating the actual variability in the population when calculating the distribution variance. To circumvent these difficulties, we estimate the actual probability distribution for both PD and TM approaches. For PD, the probability density function can be expressed analytically:

\end{multicols}

\begin{equation}
 p(S_0)= \left\{\begin{matrix}
       \frac{\tau(d\tau/2-\zeta_0)}{d(1-\tau  S_0)\ln{\big(\frac{1-s(d\tau/2)}{1-s(\zeta_0)}\big)}}-\frac{\zeta_0+d-d\tau/2}{dS_0\ln{\big(\frac{s(d\tau/2)}{s(\zeta_0+d)}\big)}} & \text{for } S_0 \in [s(\zeta_0),s(\zeta_{max})]  \\
       0 & \text{for } S_0 \not\in [s(\zeta_0),s(\zeta_{max})]
    \end{matrix}\right.
\end{equation}

\begin{multicols}{2}
However, while $p(C)$ cannot be expressed analytically for TM, it can be very efficiently estimated numerically based on the knowledge of $C(\zeta)$. Knowing $p(X)$, we estimate the spike detection efficiency (E) defined as the fraction of detected spikes for a given noise level ($n$):

\begin{equation}
E(n)=1-\text{CDF}[p(X)](N_{\sigma_n}n), 
\end{equation}
where $\text{CDF}[p(X)](n)=\int_{-\infty}^{n} p(X) dX$ is the cumulative density function, $N_{\sigma_n}=1$ for PD and $N_{\sigma_n}=\sqrt{\sum_iT_i^2}$ for TM are the noise normalization constants. This allows a more general comparison between wide-field and scanning microscopes that is detached from specific estimations of noise levels. Moreover, it provides a clear performance metric that directly relates to the fraction of detected spikes. In Fig \ref{fig:2}b we observe similar relative performances with both the PD and TM. For low sampling frequencies ($f=0.2$ and $f=0.5$) and in the low spike efficiency regime (when less than 50\% of spikes are to be detected) scanning modalities clearly outperform wide-field modalities. However, when attempting to achieve a very high detection efficiency ($>$80\%) scanning modalities severely underperform requiring a significantly higher SNR compared to wide-field. These differences disappear when a sufficient sampling frequency is achieved. To better illustrate this phenomenon, the data can be plotted along a different dimension (Fig \ref{fig:2}c) defining the relative sensitivity ($R_s$) between the two modalities for a given target spike detection efficiency as:
\begin{equation}
    R_s(E)=\frac{n(E)\big|_{\tau=0.1}}{n(E)\big|_{\tau=1}}
\end{equation}
The interpretation of $R_s$ is straightforward: it describes the ratio of the expected SNR for a given spike detection efficiency when using a wide-field over a scanning microscope when all imaging parameters are equal. Since typically only high detection efficiencies are experimentally useful, we compare the performance of the two modalities when $E$ is in the range of 50-100\% (Fig \ref{fig:2}c). In agreement  with the interpretation of Fig \ref{fig:2}b, we observe that a wide-field microscope outperforms a scanning microscope in the low sampling regime (by approximately an order of magnitude for $f=0.2$, and about twofold for $f=0.5$). However, as $f \rightarrow 1$ (which for most voltage indicators corresponds to an approximately 1 kHz frame rate) the differences become negligible. We thus observe that both PD and TM yield similar pike detection efficiencies for the equal sampling frequency ($f \approx1$), and that no systematic performance advantage comes from using a wide-field microscope. Note that this differs from our previous results obtained with the Z-score approach (Fig \ref{fig:1}h-o), presumably because the Z-score overestimates the effect of the population variance on system performance.

\begin{figure}
\begin{center}
\includegraphics[width=12cm]{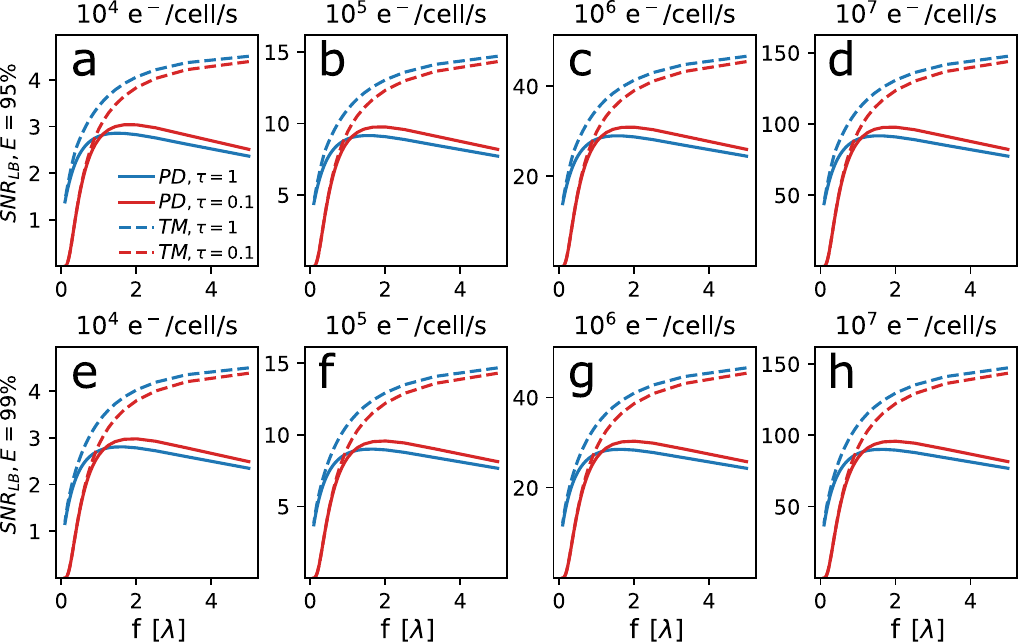} 
\caption{\label{fig:3}
(a-d) Estimated lower bound of the SNR for peak detection (PM) and template matching (TM) using scanning and wide-field systems for different number of detected photons, assuming a target $E=95\%$. (e-h) Estimated lower bound of the SNR for PD and TM using scanning and wide-field systems for different number of detected photons, assuming a target $E=99\%$.}
\end{center}
\end{figure}

Finally, our analysis framework can also be used to compare the expected lower bound for the SNR of the recorded spikes (Fig \ref{fig:3}a-h), defined as 
\begin{equation}
    \text{SNR}_{LB}(E)=\frac{n(E)}{\sigma'_n}
\end{equation}
where $\sigma'_n=\sqrt{2}\sigma_n$ for PD and $\sigma'_n=\sigma_n\sqrt{2\sum_iT_i^2}$ for TD. Making use of experimentally relevant parameters (same as for Fig \ref{fig:1}h-o), we find that the shape of the $\text{SNR}_{LB}$ does not strongly depend on the target $E$ (at least in the high photon-count regime) nor the total number of detected photons (also only in the high photon-count regime), suggesting that generalized conclusions can be drawn over a broad experimentally relevant range (see Fig \ref{fig:3}a-h). Firstly, TM outperforms PD in all instances (which was already demonstrated \cite{franke2015bayes}). Secondly, scanning microscopes outperform wide-field microscopes for high sampling frequencies, but only for PD. Thirdly, wide-field microscopes are more robust to undersampling than scanning microscopes, a conclusion that is consistent between our different analysis approaches. Fourthly, our more detailed analysis (Fig \ref{fig:2},\ref{fig:3}) shows that while wide-field does outperform scanning when implementing TM, this advantage vanishes as $f \rightarrow  1$. Finally, our lower bound SNR estimation shows that for both modalities SNR improves as sampling frequency increases; however, this improvement is subject to diminishing returns, with sampling frequencies $f>2$ yielding only incremental improvement.

\section{Conclusions}

We emphasize that this work focuses solely on the effects of temporal integration on the measured spike shapes and statistics, neglecting other differences between scanning and wide-field microscopes. For example, scanning microscopes can operate in confocal mode, yielding improved background rejection and, under certain conditions, improved SNR \cite{xiao2024large}. Similarly, there has been a growing interest in 2-photon voltage imaging \cite{villette2019ultrafast,liu2022sustained}, which, due to its nonlinear dependence on excitation power, is largely limited to a scanning configuration \cite{phil2024optical}. In either case, the sampling rate must be chosen sufficiently high to avoid performance degradation of scanning systems. This work allows for a quantitative estimation of the threshold sampling frequency ($f^*\approx\lambda$) which for most commonly used voltage sensors would fall around 1 kHz. While reducing the frame rate to 500 Hz will lead to a 2-fold degradation of the relative sensitivity for the scanning system, this might be acceptable depending on the imaging conditions. However, our work does strongly advise against using scanning systems below the 500 Hz threshold as it might lead to an explosive degradation in performance. The conclusions of this paper are directly applicable to other signal types that are transient and exponential, such as signals from calcium \cite{tian2009imaging} or glutamate indicators \cite{marvin2013optimized}; however, they are most relevant to voltage imaging, where the high frame rates required still pose significant technical challenges.

\end{multicols}

\section*{Appendix 1}

This Appendix provides explicit expressions for results obtained in this paper. 

The expected value of signal sampling is

\begin{equation}
    E[S_i(\zeta)]=\left\{\begin{smallmatrix} 
    0 & \text{for } i \leq -2\\
    \frac{1}{2}+\frac{\zeta_0}{d'}+\frac{1}{d'\lambda}[e^{-\lambda (\zeta_0+d'/2)}-1] & \text{for } i=-1\\
    \frac{1}{2}-\frac{\zeta_0}{d'}+\frac{1}{d'\lambda}[e^{-\lambda d'}-e^{-\lambda(\zeta_0+d'/2)}]-\frac{2 \sinh{(\lambda d'/2)}}{d' \lambda}[e^{-\lambda(\zeta_0+d)}-e^{-\lambda d'/2}] & \text{for } i = 0\\
    -\frac{2 \sinh{(\lambda d'/2)}}{d' \lambda}e^{-\lambda(\zeta_0+di)}[e^{-\lambda d}-1] & \text{for } i > 0\\
    \end{smallmatrix}\right.
\end{equation}
where $d'=d\tau$ and $\zeta_0$ can be calculated from

\begin{equation}
    \zeta_0=-\frac{1}{\lambda}ln\bigg[1-\frac{\tau e^{\lambda d (\tau/2-1)}}{\tau e^{\lambda d (\tau/2-1)}+\frac{\tau}{2}\sinh{(\lambda  d'/2)}}\bigg]-\frac{d'}{2}
\end{equation}

The variance can be expressed as:

\begin{equation}
    V[S_i(\zeta)]=E[S_i(\zeta)^2]-E[S_i(\zeta)]^2,
\end{equation}

were:

\begin{equation}
    E[S_i(\zeta)^2]=\left\{\begin{smallmatrix} 
    0 & \text{for } i \leq -2\\
    \frac{1}{d\tau^2}[\zeta_0+\frac{d'}{2}+\frac{1}{2\lambda}(4e^{-\lambda(\zeta_0+d'/2)}-e^{-2\lambda(\zeta_0+d'/2)}-3)] & \text{for } i=-1\\
    \frac{d'}{2}-\zeta_0+\frac{2}{\lambda}[e^{-\lambda d'}-e^{-\lambda(\zeta_0+d'/2)}]-\frac{1}{2\lambda}[e^{-2\lambda d'}-e^{-2\lambda(\zeta_0+d'/2)}]-\frac{2\sinh{(\lambda d'/2)}}{\lambda}[e^{-2\lambda(\zeta_0+d)}-e^{-\lambda d'}] & \text{for } i = 0\\
    -\frac{2}{\lambda \tau d'} \sinh{(\lambda d'/2)}^2 e^{-2 \lambda (\zeta_0+di)} [e^{-2\lambda d}-1]& \text{for } i > 0.\\
    \end{smallmatrix}\right.
\end{equation}

The covariance can be expressed as:

\begin{equation}
    Cv[S_i(\zeta),S_j(\zeta)]=\left\{\begin{smallmatrix} 
    V[S_i(\zeta)] & \text{for } i = j\\
    E[S_i(\zeta)S_j(\zeta)]-E[S_i(\zeta)]E[S_j(\zeta)] & \text{for } i \neq j,\\
    \end{smallmatrix}\right.
\end{equation}

where:

\begin{equation}
    E[S_i(\zeta)S_j(\zeta)]=\left\{\begin{smallmatrix} 
    0 & \text{for } i < -1\\
    Ae^{-\lambda d}[e^{-\lambda \zeta_0}-\frac{1}{2}e^{-\lambda(2\zeta_0+d'/2)}-\frac{1}{2}e^{\lambda d'/2}] & \text{for } i = -1, j>-1\\
    A[e^{-\lambda d'/2}(1-\frac{1}{2}e^{-\lambda d'}+\frac{1}{2}e^{-2\lambda \zeta_0})-e^{-\lambda \zeta_0}+\sinh{(\lambda d'/2)}(e^{-2\lambda(\zeta_0+d)}-e^{-\lambda d'})] & \text{for } i=0,j>0\\
   A\sinh{(\lambda d'/2)} e^{-\lambda(2\zeta_0+id)}[e^{2\lambda d}-1] & \text{for } i,j>0,\\
    \end{smallmatrix}\right.
\end{equation}
where $A=-\frac{2}{\lambda \tau d'}\sinh{(\lambda d'/2)} e^{-\lambda jd}$.

\section*{Appendix 2}

We show here that the models for point- and line-scanning systems quickly converge even for sampling values below those typically used in voltage imaging. 

The difference between point and line scanning lies in the fact that the signal integral over a longer dwell time (line scan) is substituted by several delta-like samplings (point scan) with the same average power (Fig \ref{fig:1}a). Because of that this problem is similar to error estimation when comparing analytical and numerical integration and the difference ($\delta$) between the signal recorded using a point-scan and a line-scan can be expressed as: 

\begin{equation}
    \delta =\bigg| \int_{t_0}^{t_0+\Delta t}  v(t) dt - \Delta  t v(t_0)\bigg| \leq \frac{\Delta  t^2}{2} \sup_{t \leq t \leq t_0+\Delta t} |v'(t)|,
\end{equation}
where $v'(t)$ is the time derivative of $v(t)$. For simplicity of calculations we overestimate the error by using a global supremum and calculate the error estimation relative to the maximum  value of the function

\begin{equation}
    \frac{\delta}{\max v(t)} \leq \frac{\sup_{x \in \mathbb{R}}|v'(t)|}{2  \max v(t)} \Delta t^2 = \alpha \frac{\tau^4}{f^2},
    \label{eq: A2}
\end{equation}
where $\tau \approx 0.1$ is a good approximation for an actual voltage imaging microscope, $\alpha = 0.5$ for an exponential spike. Equation \ref{eq: A2} shows that a point-scanning microscope can be reasonably approximated using a line-scanning  model  ($\delta / \max v(x)    \approx 5 \times 10^{-5} / f^2$) even at a slow (for voltage imaging) sampling rate of $f=0.1$ (where the relative error is $\leq 0.5\%$). For comparison the  difference between a wide-field and a line-scan at the same frequency is only guaranteed to be $\leq 5000\%$.

\section*{Funding}
This work was funded by the National Institutes of Health (R34NS127098).

\section*{Disclosures}
Authors declare no conflicts of interest.

\section*{Code, Data, and Materials Availability}
Code used to generate the results presented in this paper is not publicly available at this time but may be obtained from the authors upon reasonable request.



\printbibliography

\end{document}